\documentclass{aa}
\usepackage{lscape}
\usepackage{graphicx}

\usepackage{natbib}
\bibpunct{(}{)}{;}{a}{}{,} 

\usepackage{txfonts}

\newcommand{\RVstop}{20 Dec 2018}
\newcommand{\Ntransits}{two}
\newcommand{\Nstar}{HD\,1397} 
\newcommand{\Nplanet}{HD\,1397b} 
\newcommand{\NstarTOI}{TOI-120.01} 

\newcommand{\vmag}{$7.868 \pm 0.02$} 
\newcommand{\vmagshort}{7.9} 
\newcommand{\VSini}{2} 
\newcommand{\Prot}{$42.5 \pm 3.4$} 
\newcommand{\Pact}{$47.3^{+3.7}_{-4.1}$} 

\newcommand{\age}{$4.51 \pm 0.5$} 
\newcommand{\NstarTeff}{$5521\pm60$} 
\newcommand{\FeH}{$0.29\pm0.09$} 
\newcommand{\NstarRad}{$2.336^{+0.052}_{-0.059}$} 
\newcommand{\Nstarlogg}{$3.823\pm0.02$} 
\newcommand{\Nstardense}{$0.147 \pm 0.01$} 
\newcommand{\NstarMass}{$1.324^{+0.042}_{-0.046}$} 

\newcommand{\NperiodShort}{$11.54$} 
 \newcommand{\Nperiod}{\mbox{$11.5353 \pm 0.0008$}}
\newcommand{\NplanetRad}{$1.026 \pm 0.026$} 
\newcommand{\NplanetMass}{$0.415\pm 0.020$} 
\newcommand{\NplanetDense}{$0.477^{+0.043}_{-0.038}$} 
\newcommand{\Necc}{$0.251\pm0.020$}

\newcommand{\Ndist}{$79.31 \pm 0.17 $} 
\newcommand{\Av}{$0.034 \pm 0.02$} 

\newcommand{\kms}{km\,s$^{-1}$}

\newcommand{\ms}{m\,s$^{-1}$}

\newcommand{\cmss}{cm\,s$^{-2}$}

\newcommand{\gccc}{g\,cm$^{-3}$}
\newcommand{\masy}{mas\,yr$^{-1}$}
\newcommand{\mjup}{M$_{J}$}
\newcommand{\rjup}{R$_{J}$}

\newcommand{\msun}{$M_{\odot}$}
\newcommand{\rsun}{$R_{\odot}$}
\newcommand{\teff}{$T_{\rm eff}$}

\newcommand{\LSO}{La Silla Observatory}
\newcommand{\PAR}{Paranal Observatory}

\newcommand{\kepler}{{\it Kepler}}
\newcommand{\tess}{{\it TESS}}

\newcommand{\gaia}{{\it Gaia}}

\newcommand{\exofast}{{\it EXOFASTv2}}
\newcommand{\emp}{\textsc{SpecMatch-emp}}

\providecommand{\bjdtdb}{\ensuremath{\rm {BJD_{TDB}}}}

\providecommand{\teff}{\ensuremath{T_{\rm eff}}}

\providecommand{\msun}{\ensuremath{\,M_{\sun}}}
\providecommand{\rsun}{\ensuremath{\,R_{\sun}}}
\providecommand{\lsun}{\ensuremath{\,L_{\sun}}}
\providecommand{\mj}{\ensuremath{\,M_{\rm J}}}
\providecommand{\rj}{\ensuremath{\,R_{\rm J}}}

\providecommand{\fave}{\langle F \rangle}
\providecommand{\fluxcgs}{10$^9$ erg s$^{-1}$ cm$^{-2}$}
\usepackage[backref,colorlinks=true,citecolor=blue]{hyperref}

\begin{document}

   \title{A Jovian planet in an eccentric 11.5 day orbit around \object{HD~1397} discovered by \tess }



   \author{L.D.\ Nielsen \inst{1}\fnmsep\thanks{Email: Louise.Nielsen@unige.ch}
   \and
   F. Bouchy \inst{1}
   \and
   O. Turner \inst{1}
   \and
   H. Giles \inst{1}
   \and
   A. Su\'arez Mascare\~{n}o \inst{1}
   \and 
   C. Lovis \inst{1}
   \and
   M. Marmier \inst{1}
   \and
   F. Pepe \inst{1}
   \and
   D. S{\'e}gransan \inst{1}
   \and
   S. Udry \inst{1}
   \and
   J.F. Otegi \inst{1,2}
   \and
   G. Ottoni \inst{1}
   \and
   M. Stalport \inst{1}
   \and 
   G. Ricker \inst{8}
   \and
   R. Vanderspek \inst{8}
   \and
   D.W. Latham \inst{7}
   \and
   S. Seager \inst{8,12}
   \and
   J.N.\ Winn \inst{13}
   \and
   J.M.\ Jenkins \inst{14}
   \and 
   S.R. Kane \inst{3} 
   \and 
   R.A. Wittenmyer \inst{4}
   \and
   B. Bowler \inst{18}
   \and
   I. Crossfield \inst{8}
   \and
   J. Horner \inst{4}
   \and
   J. Kielkopf \inst{19}
   \and
   T. Morton \inst{20}
   \and
   P. Plavchan \inst{21}
   \and
C.G. Tinney \inst{22}
\and
Hui Zhang \inst{23}
\and 
D.J. Wright \inst{4}
\and
M.W. Mengel \inst{4}
\and
J.T. Clark \inst{4}
\and
J. Okumura \inst{4}
\and
B. Addison \inst{4}
   \and 
   D.A. Caldwell \inst{5,14} 
   \and
   S.M. Cartwright \inst{6} 
   \and
   K.A. Collins \inst{7} 
   \and
   J. Francis \inst{8}
   \and
   N. Guerrero \inst{8} 
   \and
   C.X. Huang \inst{8}
   \and
   E.C. Matthews \inst{8}
   \and
   J. Pepper \inst{9}
   \and
   M. Rose \inst{10} 
   \and
   J. Villase\~{n}or \inst{8} 
   \and
   B. Wohler  \inst{5,14} 
   \and
   K. Stassun \inst{11}
   \and
    S. Howell \inst{14}
    \and
    D. Ciardi \inst{15}
    \and
    E. Gonzales \inst{16}
    \and
    R. Matson \inst{14}
    \and
    C. Beichman \inst{15}
    \and
    J. Schlieder \inst{17}
   }
   
   \institute{Geneva Observatory, University of Geneva, Chemin des Mailettes 51, 1290 Versoix, Switzerland
   \and 
   Institute for Computational Science, University of Zurich, Winterthurerstr. 190, CH-8057 Zurich, Switzerland
   \and 
   Department of Earth Sciences, University of California, Riverside, CA 92521, USA
   \and 
   University of Southern Queensland, Centre for Astrophysics, Toowoomba, QLD 4530, Australia
   \and 
   SETI Institute, Mountain View, CA 94043, USA
   \and 
   Proto-Logic LLC, Washington, DC 20009, USA
   \and 
   Harvard-Smithsonian Center for Astrophysics, 60 Garden Street, Cambridge, MA 02138, USA
   \and 
   Department of Physics and Kavli Institute for Astrophysics and Space Research, MIT, Cambridge, MA 02139, USA
   \and 
   Department of Physics, Lehigh University, 16 Memorial Drive East, Bethlehem, PA 18015, USA
   \and 
   Leidos/NASA Ames Research Center, Moffett Field, CA 94035, USA
   \and 
   Dept. of Physics \& Astronomy, Vanderbilt University, 6301 Stevenson Center Ln., Nashville, TN 37235, USA
   \and 
   Dept. of Earth, Atmospheric, and Planetary Sciences, and Dept. of Aeronautics and Astronautics, MIT, 77 Massachusetts Avenue, Cambridge, MA 02139, USA
   \and 
   Department of Astrophysical Sciences, Princeton University, 4 Ivy Lane, Princeton, NJ 08544, USA
   \and 
   NASA Ames Research Center, Moffett Field, CA 94035, USA
   \and 
   Caltech/IPAC-NASA Exoplanet Science Institute, M/S 100-22, 770 S. Wilson Ave, Pasadena, CA 91106 USA
   \and 
   UC Santa Cruz, Department of Astronomy \& Astrophysics, 1156 High Street, Santa Cruz, CA 95064, USA
   \and 
   NASA Goddard Space Flight Center, 8800 Greenbelt Rd., Greenbelt, MD 20771, USA
   \and 
   The University of Texas at Austin, Department of Astronomy, 2515 Speedway C1400, Austin, TX 78712, USA
   \and 
   Department of Physics and Astronomy, University of Louisville, Louisville, KY 40292, USA
   \and 
   Astronomy Department, University of Florida, 211 Bryant Space Science Center, P.O. Box 112055, Gainesville, FL 32611-2055, USA
   \and 
   Department of Physics and Astronomy, George Mason University, Fairfax, VA 22030, USA
   \and 
   Exoplanetary Science at UNSW, School of Physics, UNSW Sydney, NSW, 2052, Australia
   \and 
   School of Astronomy and Space Science, Key Laboratory of Ministry of Education, Nanjing University, Nanjing, China
   }

   \date{Submitted 5 November 2018 / Accepted 18 February 2019}

 \abstract{The Transiting Exoplanet Survey Satellite \tess\ has begun a new age of exoplanet discoveries around bright host stars. 
 We present the discovery of \Nplanet\ (\NstarTOI), a giant planet in an \NperiodShort-day 
 eccentric orbit around a bright (V=\vmagshort) G-type subgiant. We estimate both host star and planetary parameters consistently using \exofast\, based on \tess\ time-series photometry of transits and radial velocity measurements from CORALIE and MINERVA-Australis. We also present high angular resolution imaging with NaCo to rule out any nearby eclipsing binaries. We find that \Nplanet\ is a Jovian planet, with a mass of \NplanetMass\ \mjup\ and a radius of \NplanetRad\ \rjup. Characterising giant planets in short-period eccentric orbits, such as \Nplanet, is important for understanding and testing theories for the formation and migration of giant planets as well as planet-star interactions.
 }


\maketitle

\section{Introduction}
Transiting exoplanets offer a unique window into exoplanetology, because we can measure both the mass and radius of the planet, and thereby place constraints on the interior structure. Atmospheric characterisation is also possible through transmission spectroscopy, thus enabling a full understanding of bulk properties and atmosphere.

After the recent end of the NASA {\it Kepler} and {\it K2} missions, the exoplanet-finding torch has truly been passed on to the Transiting Exoplanet Survey Satellite \citep[\tess\ -][]{Ricker2015}. Since the start of science operations on July 25, 2018, \tess\ has successfully delivered several hundred exoplanet candidates, and the first few planet confirmations have been reported e.g., \cite{Huang2018,Gandolfi2018,Vanderspek2018,Wang2018}.

In this paper we present the discovery of a jovian planet in an eccentric \NperiodShort-day orbit around \Nstar, a bright (V=7.9) sub-giant star from \tess\, with a mass characterisation enabled by CORALIE and MINERVA-Australis. 


\begin{table}
\centering
\caption{\label{tab:stellar} Stellar Properties for \Nstar.}
\resizebox{\columnwidth}{!}{%
	\begin{tabular}{lcc}
	\hline\hline 
	Property	&	Value	&	Source\\
	\hline
    \multicolumn{3}{l}{Other Names}\\
    2MASS ID	& J00174714-6621323	& 2MASS \\
    \multicolumn{2}{l}{\gaia\ ID DR2 ~~~~~4707634458245031552}	& \gaia \\
    TIC  ID & 394137592 & \tess \\
    TOI & \NstarTOI & \tess \\
    \\
    \multicolumn{3}{l}{Astrometric Properties}\\
    R.A.		&	00:17:47.14 		& \tess	\\
	Dec			&	-66:21:32.35	& \tess	\\
    $\mu_{{\rm R.A.}}$ (\masy) & 64.762 $\pm$0.049 & \gaia \\
	$\mu_{{\rm Dec.}}$ (\masy) & -5.055 $\pm$0.041 & \gaia\\
    Parallax  (mas) & 12.609 $\pm$0.027& \gaia\tablefootmark{$\dagger$}\\
    Distance  (pc) & \Ndist & \gaia\tablefootmark{$\dagger$}\\
    \\
    \multicolumn{3}{l}{Photometric Properties}\\
	V (mag)		&\vmag  	&Tycho \\
	B (mag)		&8.75 $\pm$0.02	&Tycho\\
    G (mag)		&	7.59 $\pm$ 0.02			&{\gaia}\\
    T (mag)	&7.14 $\pm$ 0.03				&\tess\\
    J (mag)		&6.442 $\pm$ 0.02	&2MASS\\
   	H (mag)		&6.090 $\pm$ 0.04	&2MASS\\
	K$_{\rm s}$ (mag) &5.988 $\pm$ 0.02	&2MASS\\
    W1 (mag)	&6.018 $\pm$ 0.096	&WISE\\
    W2 (mag)	&5.898 $\pm$ 0.043	&WISE\\
    W3 (mag)    &5.988 $\pm$ 0.30 & WISE\\
    W4 (mag)    &5.921 $\pm$ 0.10& WISE \\
    A$_{V}$	& \Av & Sec. \ref{sec:exofast}\\
    \\
    \multicolumn{2}{l}{Bulk Properties}& This work:\\
    \teff\,(K)    & \NstarTeff    &Sec. \ref{sec:specMatch} \& \ref{sec:exofast}\\
    log g (\cmss)& \Nstarlogg & Sec. \ref{sec:exofast}\\
    $\rho$ (\gccc)& \Nstardense & Sec. \ref{sec:exofast}\\
    $\mathrm{\left[Fe/H\right]}$  & \FeH &Sec. \ref{sec:specMatch} \& \ref{sec:exofast}\\
	{\it v}\,sin\,{\it i} (\kms)	& < \VSini	& Sec. \ref{sec:specMatch} \\
	Age	(Gyrs) & \age	&	Sec. \ref{sec:exofast}\\
    Mass ($M_{\odot}$) &  \NstarMass & Sec. \ref{sec:exofast}\\
    Radius ($R_{\odot}$) &	\NstarRad & Sec. \ref{sec:exofast}\\
	\hline
    \end{tabular}
    }
\tablefoot{Tycho \citep{Tycho}; 2MASS \citep{2MASS}; WISE \citep{WISE}; \gaia\ \citep{Gaia2018}; \tablefoottext{$\dagger$}{We have added 0.07 mas to the \gaia\ parallax, see Sec. \ref{sec:exofast} for more information.}}
\end{table}

\section{Observations}
\subsection{TESS photometry}
\Nplanet\ (TIC 394137592) was observed by TESS between 2018~Jul~25 and Sep~20, in the first of the 26 sectors of the two-year survey. The target appeared on Camera 3, CCD 1 in TESS sector 1, and was observed with both 2\,min cadence data and the 30\,min cadence Full Frame Images. 
We used the publicly available TESS sector 1 data with 2-minute cadence, supplied through the TESS Alert mechanism. They were reduced with the Science Processing Operations 
Center (SPOC) pipeline, originally developed for the \kepler\ mission at the NASA Ames Research Center \citep{Jenkins:2017,Jenkins:2016}. Two transits of \Nplanet\ were detected with signal-to-noise ratio of 81.0 \citep{jenkins:2002,KDPH_TPS2017,Twicken:2018}.  For transit modelling, we used the raw light curve based on the Pre-search Data Conditioning data product \citep{ Twicken:2010} (see section \ref{sec:exofast}). The light curve precision is 144\,ppm, averaged over one hour, consistent with the value predicted by \citet{Sullivan:2015}. 

\subsection{High resolution spectroscopy with CORALIE}
\Nstar\ was observed with the high resolution spectrograph CORALIE on the Swiss $1.2\, \mathrm{m}$ Euler telescope at \LSO~\citep{CORALIE} between 8 Sep 2018 and \RVstop. CORALIE is fed by a 2\arcsec\ fibre and has resolution $R=60,000$. In total 42 measurements were obtained. For three of them the simultaneous drift computation failed due to abnormal instrument drift, and thus these three observations are left out of the radial velocity (RV) analysis, but included in the spectral analysis.

Radial velocities and line bisector spans were calculated via cross-correlation with a G2 binary mask, using the standard CORALIE data-reduction pipeline. Given the bright magnitude of \Nstar, we are not limited by photon noise, but rather the instrumental noise floor for CORALIE at around 3~\ms. The first few measurements were used for reconnaissance, to check for a visual or spectroscopic binary.  Once we saw that the 35~m/s difference in radial velocity was consistent with the ephemerides provided by \tess\, we commenced intensive follow-up observations covering more than 3 orbits. For one orbit, we aimed at obtaining two RV points per night to check for any shorter-period planets. The RVs, along with associated properties, are given in Tab.~\ref{tab:rvs}. In Fig.~\ref{fig:rvs}, we plot the RV time series along with our result from the joint modelling analysis (see Sec. \ref{sec:exofast}).

To ensure that the RV signal does not originate from cool stellar spots or a blended eclipsing binary, we checked for correlations between the line bisector span and the RV measurements \citep{Queloz2001}. We find no evidence for a correlation. A full treatment of the line bisector and activity indicators can be found in Sec. \ref{sec:activity}.  All the CORALIE spectra were combined in the stellar rest frame into one high signal-to-noise spectrum for spectral characterisation (see Sec. \ref{sec:specMatch}).

\begin{table}
\caption{\label{tab:rvs}Radial velocities from CORALIE and MINERVA-Australis}
\centering                          
\begin{tabular}{l c c c c}        
\hline\hline                 
BJD & RV  & $\sigma_{\mathrm{RV}}$ & BIS & texp \\
(JD - 2,400,000) & (\kms) & (\ms) &  (\ms) & (s) \\
\hline                        
CORALIE &&&&\\
58369.636359 & 30.74057 &	2.98 &	2.02	&1200 \\
58375.635259 & 30.80523 &	3.19 &	-18.68	&1200 \\
58376.731468 & 30.79243 &	3.16 &	-13.34	&1200 \\
58378.891590 & 30.75477 &	3.01 &	-8.69	&1200 \\
58381.570003 & 30.73645 &	4.34 &	-15.88	&1200 \\
58384.778428 & 30.74637 &	3.06 &	-15.00	&1200 \\
58385.852894 & 30.77454 &	2.79 &	-15.03	&1200 \\
58390.669317 & 30.74431 &	3.34 &	-7.50	&1200 \\
58392.715269 & 30.73574 &	2.96 &	-11.35	&1200 \\
58394.597926 & 30.74376 &	2.90 &	-15.77	&1200 \\
58396.677112 & 30.75529 &	3.27 &	-14.57	 &900 \\
58397.640151 & 30.77955 &	2.90 &	-7.94	&1800 \\
58398.658175 & 30.79881 &	2.76 &	-21.73	&1800 \\
58401.599576 & 30.76664 &	2.85 &	-11.78	&1200 \\
58404.559687 & 30.75173 &	3.02 &	-9.99	&1800 \\
58404.757294 & 30.74503 &	2.94 &	-8.09	&1800 \\
58405.723627 & 30.74952 &	4.29 &	-15.41	&1800 \\
58406.557883 & 30.75675 &	3.14 &	-1.61	&1800 \\
58406.739518 & 30.75340 &	3.71 &	-3.03	&1800 \\
58407.507920 & 30.76371 &	3.77 &	-4.10	&1800 \\
58407.630310 & 30.75709 &	2.76 &	-8.12	&1800 \\
58409.639926 & 30.79663 &	2.96 &	-11.08	&1200 \\
58410.510996 & 30.81433 &	2.94 &	-12.63	&1200 \\
58410.680271 & 30.80776 &	3.42 &	-2.82	&1200 \\
58411.510682 & 30.80878 &	3.19 &	-15.33	&1200 \\
58411.634253 & 30.81381 &	2.87 &	 3.76	&1200 \\
58413.500366 & 30.78242 &	3.22 &	-1.77	&1200 \\
58413.688169 & 30.78032 &	3.36 &	 10.84	&1200 \\
58416.500994 & 30.75439 &	3.34 &	-10.27	&1200 \\
58418.566944 & 30.76147 &	2.91 & 	-8.00   &1200 \\
58425.588090 & 30.75641 &	2.92 & -18.85	&1200 \\		
58427.578936 & 30.74277 &	2.99 & -8.93	&1200 \\		
58429.596759 & 30.75295 &	3.04 & -3.14	&1200 \\		
58432.634725 & 30.79144 &	3.15 & -6.97	&1200 \\		
58447.563936 & 30.77588 &	2.97 & -12.19	&1200 \\		
58459.595630 & 30.76749 &	3.29 & -6.27	&1200 \\		
58461.628231 & 30.74656 &	3.07 & -16.10	&1200 \\		
58465.553745 & 30.76210 &	3.05 & -9.64	&1200 \\		
58472.558140 &30.74762	&	2.93 & -14.43	&1200 \\	
\hline
MINERVA-Australis &&&&\\
58370.95008	& 30.74760 &	5.00	& -	&1200 \\
58370.95762	& 30.75290 &	5.00	& -	&1200 \\
58370.96516	& 30.74854 &	5.00	& -	&1200 \\
58370.97273	& 30.74265 &	5.00	& -	&1200 \\
58370.98027	& 30.75446 &	5.00	& -	&1200 \\
58372.02456	& 30.74646 &	5.00	& -	&1200 \\
58372.03213	& 30.74484 &	5.00	& -	&1200 \\
58372.03966	& 30.74235 &	5.00	& -	&1200 \\
58372.04723	& 30.76362 &	5.00	& -	&1200 \\
58372.05478	& 30.74624 &	5.00	& -	&1200 \\
58376.98719	& 30.80310 &	5.00	& -	&1200 \\
58377.00169	& 30.81319 &	5.00	& -	&1200 \\
58377.01618	& 30.80060 &	5.00	& -	&1200 \\
58381.95007	& 30.75478 &	5.00	& -	&1200 \\
58381.99081	& 30.74806 &	5.00	& -	&1200 \\
58382.00530	& 30.75384 &	5.00	& -	&1200 \\
58385.03784	& 30.76661 &	5.00	& -	&1200 \\
58385.06919	& 30.76483 &	5.00	& -	&1200 \\
58385.08369	& 30.76891 &	5.00	& -	&1200 \\
58393.91861	& 30.74899 &	5.00	& -	&1200 \\
58393.93313	& 30.74121 &	5.00	& -	&1200 \\
58393.94762	& 30.74176 &	5.00	& -	&1200 \\
\hline                                 
\end{tabular}
\end{table}

\subsection{High resolution spectroscopy with MINERVA-Australis}
\Nstar\ was observed with the MINiature Exoplanet Radial Velocity Array (MINERVA-Australis) located in Queensland, Australia on 7 nights between 2018 Sep 9 and 2018 Oct 2.  MINERVA-Australis (Addison et al. 2019, PASP, submitted) is an array of 0.7m telescopes feeding a stabilised $R=80,000$ spectrograph functionally identical to the US-based MINERVA array \citep{swift15}.  At this writing, MINERVA-Australis hosts a single 0.7m telescope (with a further four to be installed in 2019), feeding starlight via a 2.2\arcsec\ fibre into the spectrograph (Wright et al. 2019, in prep).  Wavelength calibration is achieved by a simultaneous ThAr lamp calibration fibre, and radial velocities are obtained by an implementation of standard mean-spectrum least-squares matching procedures \citep{Anglada12}. A total of 25 individual measurements were obtained; data from 16 Sep were excluded from the analysis due to extremely poor observing conditions.

\subsection{High angular resolution imaging with VLT/NaCo}
\begin{figure} 
  \centering
  \includegraphics[width=\columnwidth,trim={0cm 0cm 0cm 0cm},clip]{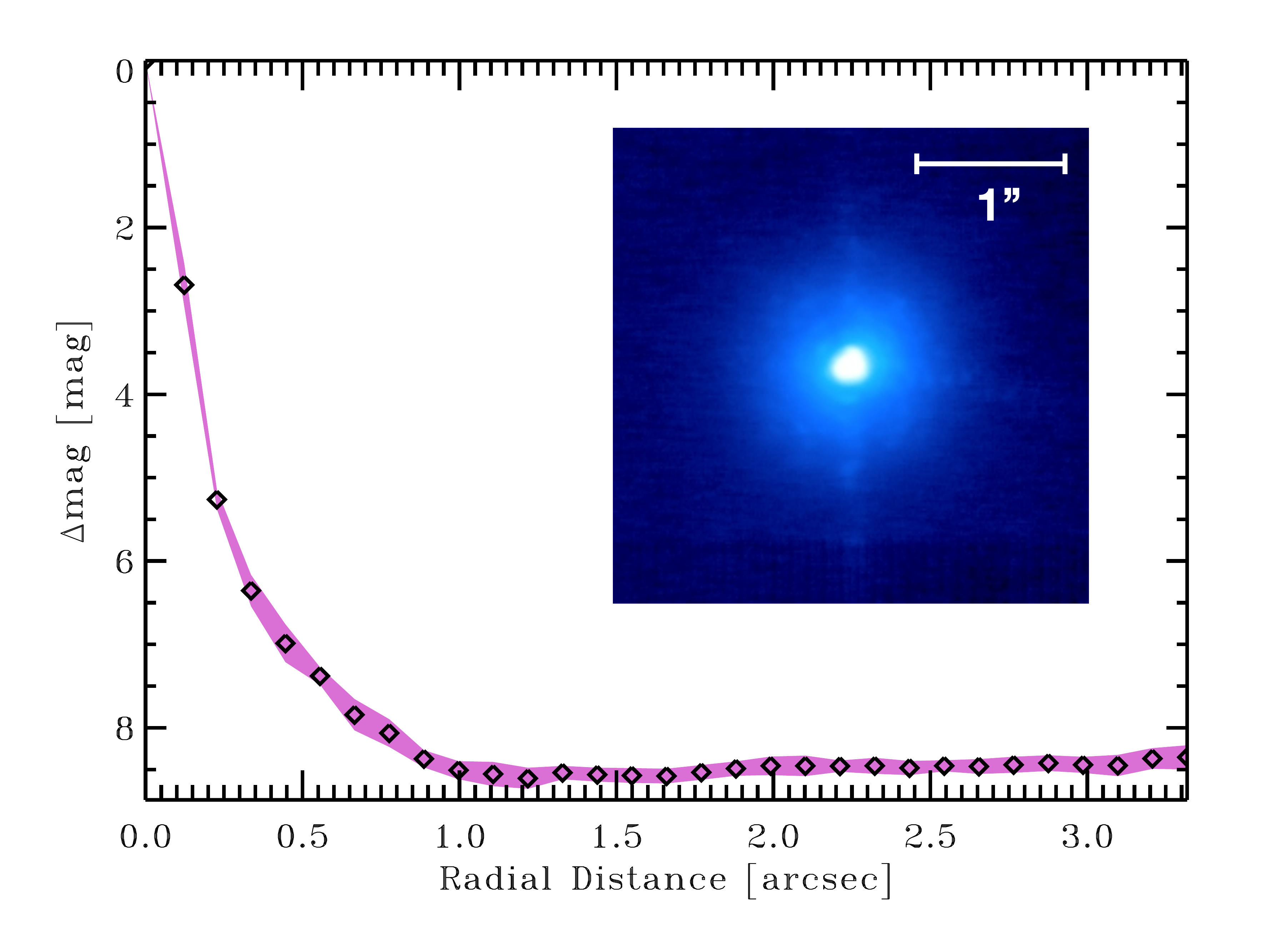}
      \caption{Sensitivity curve from NaCo AO imaging at Br-$\gamma$. The inset image is 3 arcseconds on either side.}
         \label{fig:naco}
\end{figure}

\Nstar\ was observed with NaCo \citep{Rousset2003,Lenzen2003} on the UT4 telescope of the Very Large Telescope at \PAR, on the night of 24 October 2018. A total of 54 exposures of 4 seconds each were collected using the narrowband Br-$\gamma$ filter ($\lambda_c = 2.166 ~\mathrm{\mu}$m). The target was centred within the upper left quadrant of the detector, and the telescope dithered by 2\arcsec after every second integration, with a sub-field of 4\arcsec centred on the target covered by all dithers. Images were dark subtracted and flat fielded, and the dithered images used to remove the sky background. The images were then aligned and median combined. No additional sources were identified anywhere within the final mosaiced image, which extends to at least 6 arcseconds from the target in every direction.  The target appears to be single, to the limit of the imaging resolution. To determine the sensitivity of the final combined images, simulated PSF images were inserted into the data and recovered with standard aperture photometry. A sensitivity curve is shown in Fig. \ref{fig:naco} along with the high angular resolution image of the target.

\section{Analysis \& results}

\subsection{Stellar parameters through spectral analysis} \label{sec:specMatch}
The 42 CORALIE spectra were stacked in the stellar rest frame, weighted by the individual SNR to create one high quality spectrum. Bulk stellar parameters were derived using \emp\ \citep{specmatch}, which matches the input spectra to a vast library of stars with well-determined parameters derived with a variety of independent methods, e.g., interferometry, optical and NIR photometry, asteroseismology, and LTE analysis of high-resolution optical spectra. We used the spectral region around the Mgb triplet (5100 - 5340 {\AA}) to match our spectrum to the library spectra through $\chi^2$ minimisation. A weighted linear combination of the five best matching spectra were used to extract \teff, $\log g$ and [Fe/H]. 

We adopted \teff, and [Fe/H] obtained for the stacked spectrum as priors for the global modelling. The final stellar bulk parameters, including mass and radius, are modelled jointly with the planetary parameters within \exofast\ as described in Sec. \ref{sec:exofast}. This method utilises the transit light curve, SED fitting of archival broad band photometry, and the \gaia\ parallax along with stellar tracks and isochrones. We obtain stellar radius of \NstarRad\ \rsun\ and mass \NstarMass\ \msun, as summarised in Tab. \ref{tab:stellar}.

The projected rotational velocity of the star, $v \sin i$, was computed using the calibration between $v \sin i$ and the width of the CCF from \cite{Santos2002} for CORALIE. The formal result was smaller than what can be resolved by CORALIE, and we can therefore only establish an upper limit of \VSini\ km/s.

\subsection{Stellar rotation and activity} \label{sec:activity}
To constrain the rotation period and overall stellar variability of the star we computed the time series of several spectroscopic line indicators from the CORALIE spectra. We based our analysis on the Mount Wilson Ca II H\&K S-index \citep{Noyes1984, Lovis2011}, the H$\alpha$ index \citep{GomesdaSilva2011}, He I index \citep{Boisse2009}, NA I index \citep{Diaz2007} and TiO index \citep{Azizi2018}.  We compute the power spectrum of the different time series using a generalised Lomb-Scargle periodogram~\citep{Zechmeister2009}. We then fit a double sinusoidal model at the detected period, and the first harmonic of the period, whenever there is a significant signal, in the similar way as ~\citet{Masca2017}, only this time adding a jitter parameter to the fit. 

\begin{figure}
   \centering
   \includegraphics[width=\columnwidth]{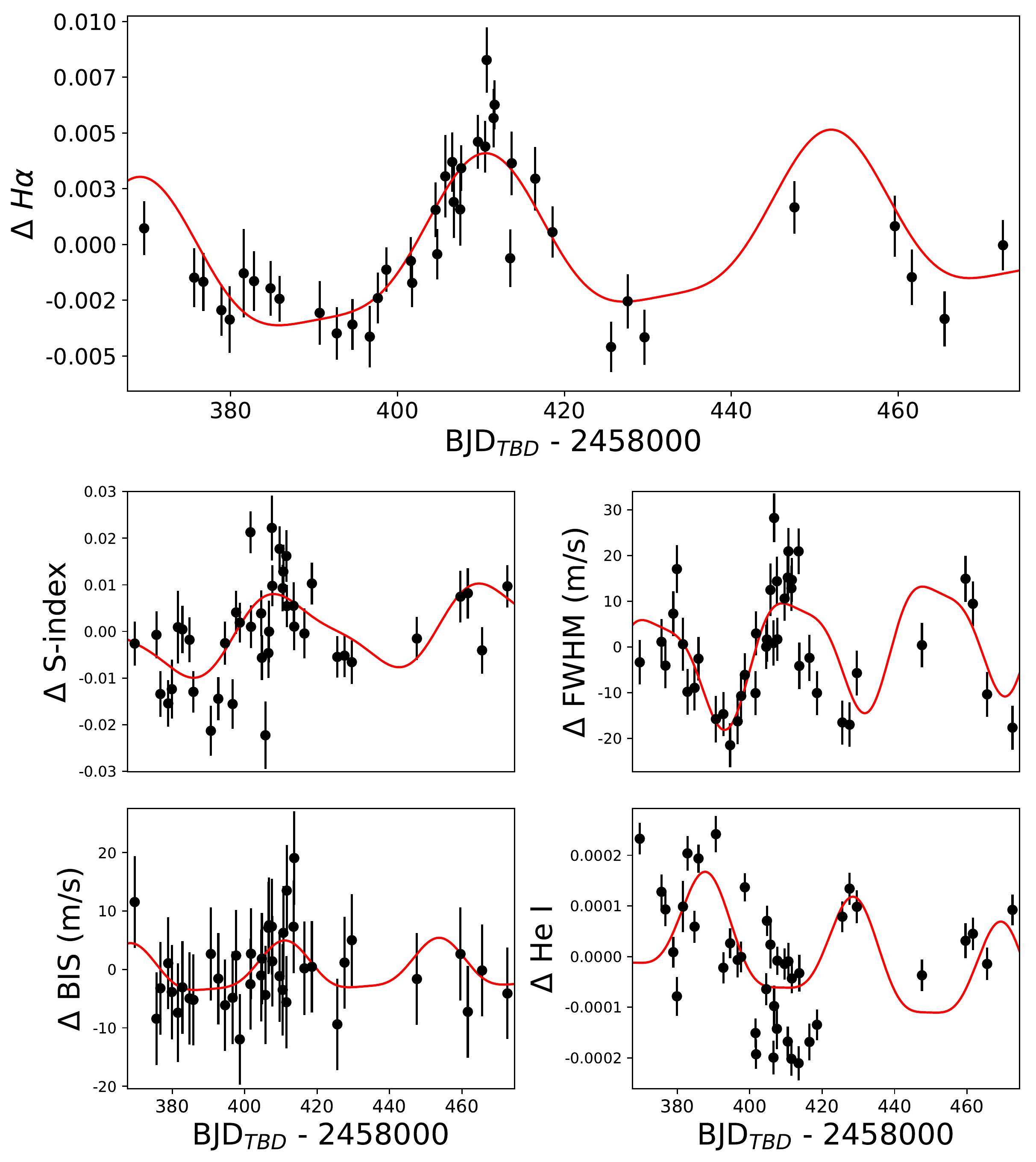}
      \caption{\label{fig:activity} Time series of the H$\alpha$ index  (top panel), Ca II H\&K S-index (mid left), FWHM (mid right), bisector span (bottom left) and He I index (bottom right) for the CORALIE spectra. The red lines show the best fit to the data.}
\end{figure}

We found evidence of variability on timescales compatible with stellar rotation. The bisector span,  FWHM of the CCF, Ca II  H\&K S-index, H$\alpha$ index and  He~I index show detections of periodicities in the range of 40-50 days (see Fig.~\ref{fig:activity}). We find no significant evidence of periodic variability in the Na I and TiO time series. This suggests the possibility of a rotation period of \Prot\ days, which would be reasonable for an evolved star, and would indicate an inclination of the rotation axis smaller than 46$^{\circ}$, considering the small $v \sin i$. This measurement should be taken with caution, as the baseline of our observations only cover two rotational periods.


We find evidence that the stellar activity is affecting the RVs. When fitting a single \NperiodShort\ day Keplerian orbit to the CORALIE RVs, the residuals correlate moderately with the H$\alpha$ index (r $\sim$  0.49). Detrending the RVs by subtracting a linear model between the two quantities does not affect the planetary detection, but reduces the planetary mass and orbital eccentricity slightly. This is taken into account in the joint modelling, as described in Sec. \ref{sec:exofast}.

\subsection{Joint modelling with \exofast} \label{sec:exofast}
Simultaneous and self-consistent fitting of the CORALIE and MINERVA-Australis radial velocities, \tess\ transit light curves and stellar parameters was performed using \exofast\ \citep[see][for a full description]{Exofastv2, Exofast}. This joint model enforces global consistency of the stellar and planetary properties with all of the available data. The parameter space is explored with a differential evolution Markov Chain method through 50,000 steps and 16 independent chains. \exofast\ has build-in diagnostics for checking how well the chains are mixing through Gelman-Rubin statistic \citep{Gelman2003} as proposed by \cite{ford2006}. In our case the Gelman-Rubin statistic indicated that all fitted parameters were well mixed. In total 31 free parameters are fitted, for which four ($\log g$, $\mathrm{[Fe/H]}$, $P_{rot}$ and the parallax) have constraining Gaussian priors, described in this section.

\begin{figure}
   \centering
      \includegraphics[width=\columnwidth]{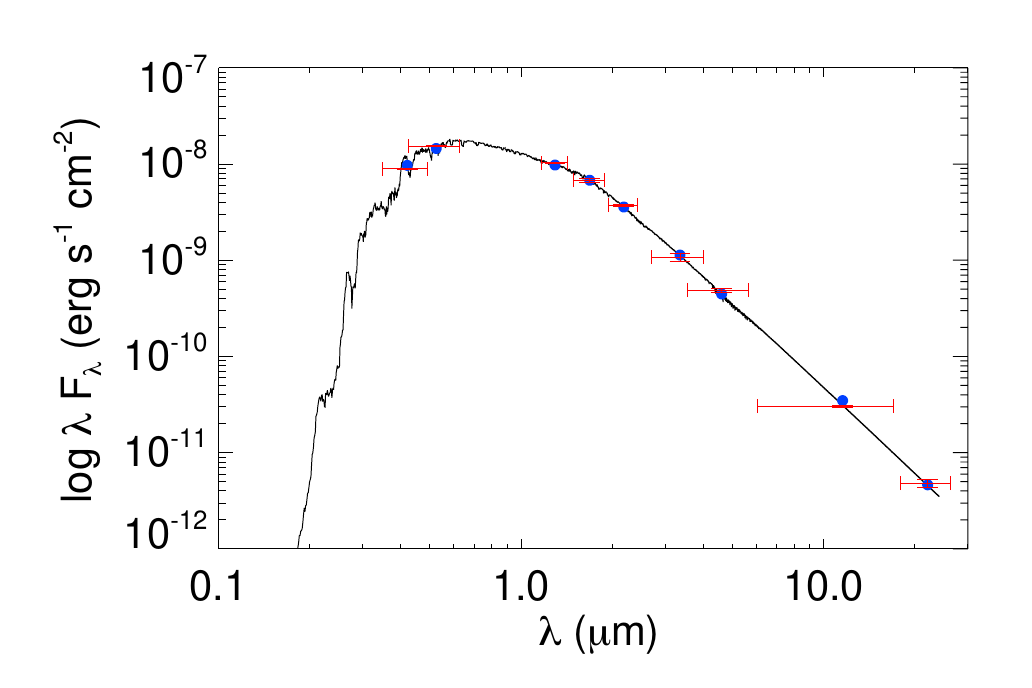}
      \caption{SED fitting of \Nstar\ as produced by \exofast. The sources of the broad band photometry is detailed in Tab. \ref{tab:stellar}.}
         \label{fig:SED}
\end{figure}

The parameters $\log g$ and $\mathrm{[Fe/H]}$ obtained from the spectroscopic analysis were used as Gaussian priors. Broadband photometry, the \gaia\ DR2 parallax and an upper limit on the V-band extinction from \cite{Schlegel1998} and \cite{Schlafly} were also used as input, to model the stellar properties consistently with the planet parameters through SED fitting. The \gaia\ DR2 parallax is a powerful parameter for computing stellar properties. For completeness, we adjusted the parallax by $+0.07$~mas based on the systematic offsets reported by \citet{StassunTorres2018} and by \citet{Zinn2018}. Both studies find the \gaia\ DR2 parallaxes to be too small by 0.06--0.08~mas from benchmark eclipsing binary and asteroseismic stars within $\sim$1~kpc. For \Nstar\ the offset represents only $\sim$0.5\% error on the parallax. However, as our aim is to achieve a precision on the radius of a few percent or better, this 0.5\% systematic error is not entirely negligible. Within \exofast\ we invoked the Mesa Isochrones and Stellar Tracks \citep[MIST][]{Mist0,Mist1} to model the star. We compared results using the Torres radius-mass relationship \citep{Torres} and YY-isochrones \citep{YY}, which showed no significant discrepancies in stellar, nor planetary, properties. The final SED fit is shown in Fig.~\ref{fig:SED}.

As demonstrated in Sec. \ref{sec:activity}, \Nstar\ is an active star which affects the RVs. To correct for the stellar activity we model the stellar rotation as an additional Keplerian with a Gaussian prior on the period from the analysis of the spectral indicators ($P_{rot} =$\Prot). A more ideal solution could be to detrend the RVs directly with an activity indicator, e.g. the H$\alpha$ index, but this is currently not possible within \exofast. Further more, we only have derived activity indicators for the CORALIE data, and would thus have to interpolate those in order to detrend the MINERVA-Australis data. We have performed an independent analysis of the CORALIE data detrended against the H$\alpha$ index, and obtained results for the planetary parameters which were consistent with the \exofast\ global model with two Keplerians. When modelling two Keplerians in \exofast, we obtain a period of \Pact\ days for the stellar activity cycle.

\begin{figure} 
   \centering   
  \includegraphics[width=\columnwidth,trim={2.0cm 12.5cm 9cm 8cm},clip]{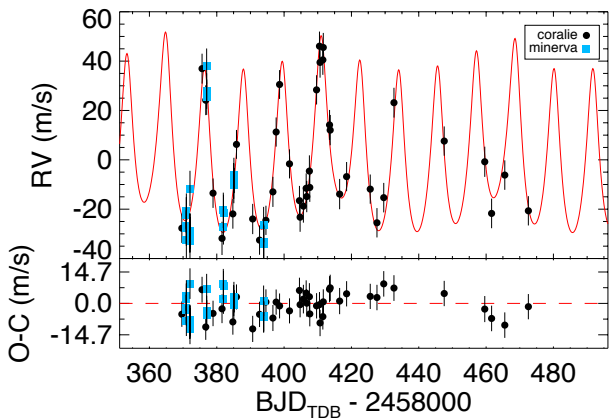}
      \caption{Time series of the CORALIE and MINERVA-Australis RVs with the \exofast\ two Keplerian model over-plotted in red.}
         \label{fig:rvs}
\end{figure}

The joint analysis shows clear evidence of a \NplanetMass\,\mjup\ and \NplanetRad\,\rjup\ planet in an eccentric orbit ($e$ = \Necc). The adopted solution with two Keplerians is illustrated in Fig.~\ref{fig:rvs} where the RV time series are plotted along with the best-fitting model.

\begin{figure}
   \centering
   \includegraphics[width=\columnwidth,trim={2.0cm 12.5cm 9cm 8cm},clip]{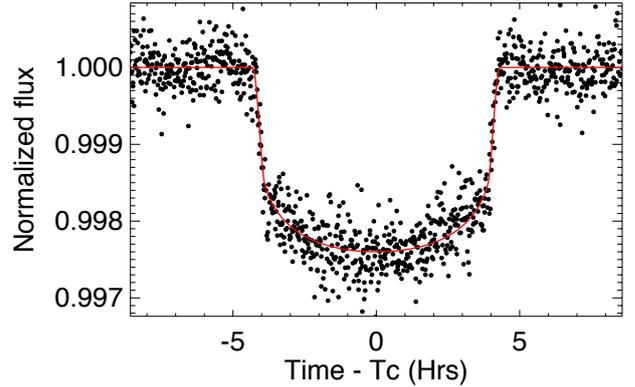}
      \caption{\label{fig:transit}The \Ntransits~ TESS transits phase-folded with the model from \exofast~ over plotted.}
\end{figure}

The period of \Nperiod\,days is well constrained, even with only \Ntransits\ \tess\ transits.
A full summary of the results from the joint modelling can be seen in Tab.~\ref{tab:stellar} and ~\ref{tab:HD1397Drift}, listing the stellar and planetary properties, respectively. The fit to the transit light curves can be seen in Fig. \ref{fig:transit}, where the \Ntransits\ \tess\ transits have been phase-folded along with the final model from \exofast.

The host star is a slightly evolved G-type sub-giant with radius \NstarRad\ \rsun\ and mass \NstarMass\ \msun. The age of the system is estimated to be \age\ Gyr which is consistent with the evolutionary stage of a star with this mass. \Nplanet\ has a mean density of \NplanetDense\,\gccc, making it a low-density exoplanet. 

\section{Discussion \& conclusion}
We present the discovery of a hot Jupiter in an \NperiodShort\ day orbit with eccentricity of \Necc, around a bright, V=7.8, G-type sub-giant\footnote{The independent study conducted by \cite{Brahm2018}, submitted on the same day as this work, also presents the discovery of \Nplanet.}. 
Planetary and stellar parameters were modelled jointly using both \tess\ transit light curves and CORALIE and MINERVA-Australis radial velocities to ensure consistent results. Through high angular resolution imaging we have also ruled out the possibility of nearby stellar companions. \Nplanet\ will, along with all the coming giant planets from \tess, help our understanding of giant planet formation, evolution and dynamics. 

\Nplanet~ is unusual in that it is a close-in planet orbiting a somewhat evolved intermediate-mass star. Nearly two decades of radial velocity surveys for planets orbiting such stars have yielded only a handful of planets with $a<0.5$ au. The majority of those surveys have targeted subgiants or low-luminosity giants \citep[e.g.][]{frink01,johnson10,jones11,witt11}, which have not yet expanded to engulf such planets \citep{kunitomo11}. Statistical analyses from these surveys have so far generally agreed that giant planets are rare within 0.5 au \citep{bowler10}, and that the overall occurrence rate of planets is positively correlated with both host-star mass and metallicity \citep{jones14,reffert15,witt17}. Tidal effects for eccentric planets orbiting giant stars result in significant tidal decay of the orbits, resulting in predictions regarding when the planet may merge with the host star \citep{wittenmyer2017}. In the case of \Nplanet, the periastron passage of the planet bring it within $\sim$8 stellar radii of the host star.

The source of \Nplanet's eccentricity adds further interest to this system.
It has been shown that sparse sampling and noisy RV data can cause circular double-planet systems to be misinterpreted as single eccentric planets \citep[e.g.][]{witt13,trifonov17,boisvert18}. In this case however, the transit detection gives assurance that the radial velocity variations are indeed due to a single planet on an eccentric orbit.  There are numerous possible origins for the nonzero orbital eccentricity, including angular momentum transfer from an additional planet in the system \citep{kane2014a}, prior planet-planet scattering events \citep{ford2008}, or the close passage of a stellar binary companion \citep{halbwachs2005,kane2014b}. Examples of the latter scenario include the extreme eccentricity cases of HD~80606b \citep{liu2018}, HD~4113b \citep{cheetham2018}, and HD~20782b \citep{kane2016}. At the present time, our data show no significant evidence for the presence of a second planet, nor is there observational evidence for a bound stellar companion to the host star.

Thus a likely scenario for the eccentricity of the orbit is perturbations from the star--planet system during the evolution of the star off the main sequence.
For example, the study by \citet{grunblatt2018} of eccentric planetary orbits associated with evolved stars concluded that the difference between tidal circularization time scales and the orbital evolution that occurs during host star mass loss may account for the observed eccentricity distribution around giant stars \citep{vl09,villaver14,veras16}. For \Nplanet, as a well-characterised planet transiting a bright star, further radial velocity measurements will either strengthen this explanation or possibly reveal evidence of additional bodies in this fascinating system.  

\begin{figure}
   \centering
   \includegraphics[width=\columnwidth]{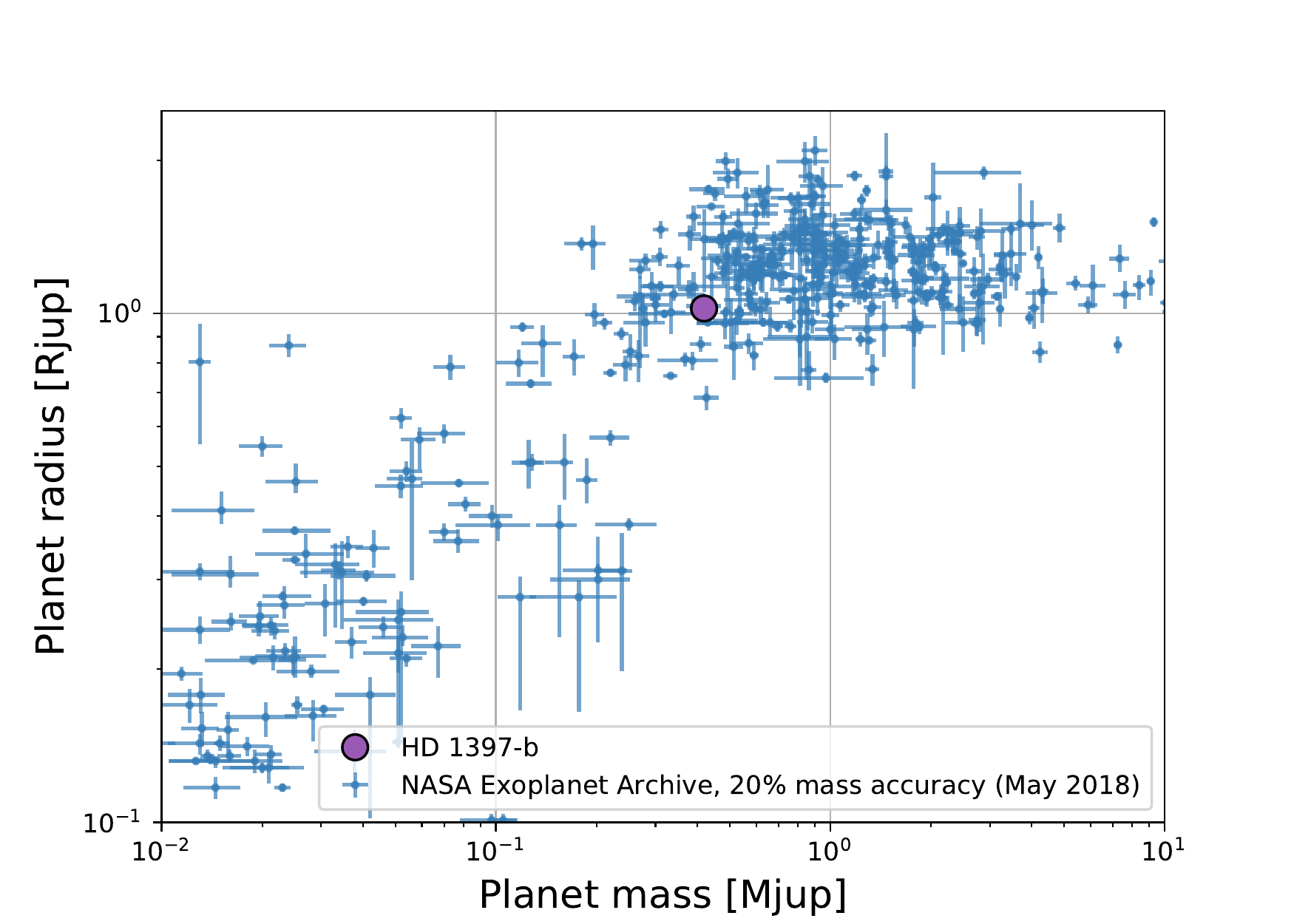}
      \caption{\label{fig:RadMass}Mass and Radius of the known population of Exoplanets with \Nplanet\ over-plotted. Only Planets with mass known to 20\% or better are included.}
\end{figure}

Jovian planets around stars that are moving up the sub-giant and red-giant branches have been proposed to be key elements in testing the theories explaining the hot Jupiter radius anomaly, by probing the possibility of late stage re-inflation \citep{LopezFortney2016,Grunblatt2017}. If the low density nature observed in hot Jupiters is caused by a fraction of incident flux deposited in the planet interior and thus inflating the planet, we expect gas giants orbiting post-main-sequence stars to be more inflated than a comparative population around dwarf stars. With its low density ($\rho_{p} =$\NplanetDense\,\gccc), \Nplanet\ is an interesting planet for testing this hypothesis. Adding more well-characterised, transiting, giant planets around sub-giant and giant stars to the known population will help further future studies explaining the hot Jupiter anomaly. Furthermore, \Nplanet's relative low mass of \NplanetMass\,\mjup, puts it on the limit what can be destroyed by Roche-lobe overflow and tidal in-spiralling, two mechanisms that have been forward to explain the evolution of gas giants to rocky super Earths, such as CoRoT-7b \citep{Jackson2010}. Figure~\ref{fig:RadMass} compares \Nplanet\ with the known population of well characterised giant exoplanets, including the transition between ice- and gas-giant planets.

\Nplanet\ is an interesting target for high resolution transmission spectroscopy to detect e.g. sodium \citep[as done by ][]{Wyttenbach2017} as well as low resolution, space-based spectroscopy searching for helium \citep{Spake2018}. It has an estimated scale height of 650 km, corresponding to a transmission signal of 35 ppm. Recently \cite{anderson17} used a metric to compare the detectability of an atmospheric signal accounting for the transmission signal and K-band stellar flux. By this metric, WASP-43\,b has a value of 74 and is notable as having had an atmospheric water abundance measured by \cite{Kreidberg14}. WASP-107\,b is used as the fiducial measure of 1000 and \Nplanet\ has a value of 490 suggesting atmospheric investigation would be viable. However, ground-based transit spectroscopy might prove difficult due to the long duration of the transit ($\sim$8.6 hr), making it hard to get enough baseline observations before and after transits.

A detection of a secondary eclipse would allow us to put further constraints on the eccentricity of the system and get a direct measure of the planet day-side temperature \citep{Charbonneau2003} and potentially begin to map the upper cloud deck temperature distribution \citep{Williams06,Knutson2007}. The eclipse time computed on the basis of the orbital parameters found in this study is $2458328.12 \pm 0.14$ in $\mathrm{BJD_{TDB}}$. The duration is expected to be $0.365 \pm 0.0014$ days. We searched the TESS light curves for a secondary eclipse, and can exclude a depth of 100 ppm in the TESS band. This is consistent with secondary eclipse depths estimated by the \exofast\ analysis in {\it Spitzer}'s $3.6-$ and $4.5- \mu$m bands of $84.5\pm 2.7$ ppm and $127.3\pm 3.5$ ppm respectively.


\begin{acknowledgements}
We  thank  the  Swiss  National  Science  Foundation  (SNSF) and the Geneva University for their continuous support to our planet search programs. This work has been in particular carried out in the frame of the National Centre for Competence in Research ‘PlanetS’ supported by the Swiss National Science Foundation (SNSF). 
This publication makes use of The Data \& Analysis Center for Exoplanets (DACE), which is a facility based at the University of Geneva (CH) dedicated to extrasolar planets data visualisation, exchange and analysis. DACE is a platform of the Swiss National Centre of Competence in Research (NCCR) PlanetS, federating the Swiss expertise in Exoplanet research. The DACE platform is available at \url{https://dace.unige.ch}. 
We acknowledge the use of TESS Alert data, which is currently in a beta test phase, from the TESS Science Office and at the TESS Science Processing Operations Center. Funding for the TESS mission is provided by NASA’s Science Mission directorate. 
This work has made use of data from the European Space Agency (ESA) mission
\gaia\ (\url{https://www.cosmos.esa.int/gaia}), processed by the \gaia\
Data Processing and Analysis Consortium (DPAC,
\url{https://www.cosmos.esa.int/web/gaia/dpac/consortium}). Funding for the DPAC has been provided by national institutions, in particular the institutions participating in the \gaia\ Multilateral Agreement.
This study was in part based on observations collected at the European Southern Observatory under ESO programme 0102.C-0503(A).
MINERVA-Australis is supported by Australian Research Council LIEF Grant LE160100001, Discovery Grant DP180100972, Mount Cuba Astronomical Foundation, and institutional partners University of Southern Queensland, MIT, Nanjing University, George Mason University, University of Louisville, University of California Riverside, University of Florida, and University of Texas at Austin.

\end{acknowledgements}

\bibliographystyle{aa} 
\bibliography{HD1397}
\begin{table*}
\caption{\label{tab:HD1397Drift} Median values and 68\% confidence intervals for HD~1397 fitted with \exofast, including two keplearians, of which the 42-day period one tracks stellar activity.}
\centering
\begin{tabular}{lccc}
\hline 
~~~Parameters & Description and units & \multicolumn{2}{c}{Values} \\ \hline
\smallskip\\\multicolumn{2}{l}{Planetary Parameters:}&b\smallskip&\\
~~~~$P$\dotfill &Period (days)\dotfill &$11.53533^{+0.00079}_{-0.00080}$&\\
~~~~$R_P$\dotfill &Radius (\rj)\dotfill &$1.026^{+0.025}_{-0.027}$&\\
~~~~$M_P$\dotfill &Mass (\mj)\dotfill &$0.415\pm0.020$&\\
~~~~$\rho_P$\dotfill &Density (cgs)\dotfill &$0.477^{+0.043}_{-0.038}$&\\
~~~~$logg_P$\dotfill &Surface gravity \dotfill &$2.991\pm0.029$&\\
~~~~$T_C$\dotfill &Time of conjunction (\bjdtdb)\dotfill &$2458332.08261\pm0.00060$&\\
~~~~$a$\dotfill &Semi-major axis (AU)\dotfill &$0.1097^{+0.0011}_{-0.0013}$&\\
~~~~$i$\dotfill &Inclination (Degrees)\dotfill &$88.99^{+0.67}_{-0.73}$&\\
~~~~$e$\dotfill &Eccentricity \dotfill &$0.251^{+0.020}_{-0.019}$&\\
~~~~$\omega_*$\dotfill &Argument of Periastron (Degrees)\dotfill &$2.1^{+6.1}_{-7.0}$&\\
~~~~$T_{eq}$\dotfill &Equilibrium temperature (K)\dotfill &$1228.3^{+10.}_{-9.9}$&\\
~~~~$K$\dotfill &RV semi-amplitude (m/s)\dotfill &$32.0\pm1.4$&\\
~~~~$R_P/R_*$\dotfill &Radius of planet in stellar radii \dotfill &$0.04513^{+0.00030}_{-0.00025}$&\\
~~~~$a/R_*$\dotfill &Semi-major axis in stellar radii \dotfill &$10.10^{+0.24}_{-0.22}$&\\
~~~~$\delta$\dotfill &Transit depth (fraction)\dotfill &$0.002037^{+0.000027}_{-0.000023}$&\\
~~~~$\tau$\dotfill &Ingress/egress transit duration (days)\dotfill &$0.01592^{+0.0011}_{-0.00042}$&\\
~~~~$T_{14}$\dotfill &Total transit duration (days)\dotfill &$0.3584^{+0.0014}_{-0.0013}$&\\
~~~~$b$\dotfill &Transit Impact parameter \dotfill &$0.17^{+0.12}_{-0.11}$&\\
~~~~$\Theta$\dotfill &Safronov Number \dotfill &$0.0671^{+0.0035}_{-0.0033}$&\\
~~~~$\fave$\dotfill &Incident Flux (\fluxcgs)\dotfill &$0.485^{+0.017}_{-0.016}$&\\
~~~~$d/R_*$\dotfill &Separation at mid transit \dotfill &$9.39^{+0.47}_{-0.42}$&\\
\smallskip\\\multicolumn{2}{l}{Stellar Parameters (See Tab. \ref{tab:stellar} for additional results:)}&\smallskip\\
~~~~$L_*$\dotfill & Luminosity (\lsun)\dotfill & $4.57^{+0.15}_{-0.16}$ &\\
~~~~$[{\rm Fe/H}]_{0}$\dotfill &Initial Metallicity \dotfill &$0.265\pm0.088$\\
~~~~$EEP$\dotfill &Equal Evolutionary Point \dotfill &$461.9^{+4.4}_{-4.1}$\\
~~~~$\sigma_{SED}$\dotfill &SED photometry error scaling \dotfill &$3.62^{+1.5}_{-0.88}$\\
\smallskip\\\multicolumn{2}{l}{Stellar activity keplearian:}&\smallskip&\\
~~~~$P_{activity}$\dotfill &Period (days)\dotfill &$47.3^{+3.7}_{-4.1}$&\\
~~~~$T_{C,activity}$\dotfill &Time of conjunction (\bjdtdb)\dotfill & $2458327.1^{+14}_{-9.1}$ \\
~~~~$e_{activity}$\dotfill &Eccentricity \dotfill & $0.28^{+0.14}_{-0.17}$\\
~~~~$\omega_{*,activity}$\dotfill &Argument of Periastron (Degrees)\dotfill & $-18^{+62}_{-83}$\\
~~~~$K_{activity}$\dotfill &RV semi-amplitude (m/s)\dotfill & $5.6^{+2.0}_{-2.3}$\\
\smallskip\\\multicolumn{2}{l}{Wavelength and Transit Parameters }&\tess\\
~~~~$u_{1}$\dotfill &linear limb-darkening coeff \dotfill &$0.340\pm0.031$\\
~~~~$u_{2}$\dotfill &quadratic limb-darkening coeff \dotfill &$0.280^{+0.044}_{-0.045}$\\
~~~~$\sigma^{2}$\dotfill &Added Variance \dotfill &$ 2.50 \cdot 10^{-8} \pm 0.16 \cdot 10^{-8}$\\
~~~~$F_0$\dotfill &Baseline flux \dotfill &$1.0000416^{+0.0000042}_{-0.0000043}$\\
\smallskip\\\multicolumn{2}{l}{Telescope Parameters (RV):}& CORALIE & MINERVA-Australis\\
~~~~$\gamma_{\rm rel}$\dotfill &Relative RV Offset (m/s)\dotfill &$30767.8\pm1.2$&$30771.8^{+2.3}_{-2.2}$\\
~~~~$\sigma_J$\dotfill &RV Jitter (m/s)\dotfill &$5.9^{+1.3}_{-1.1}$&$5.1\pm1.8$\\
\smallskip \\
\hline
\end{tabular}
\end{table*}

\end{document}